\documentclass[aps,reprint,showpacs,amsmath,amssymb,prl,superscriptaddress]{revtex4-1}

\usepackage{graphicx}
\usepackage{dcolumn}
\usepackage{bm}
\usepackage{epsf}
\usepackage{amsmath}
\usepackage{amssymb}
\usepackage{color}

\newcommand{\la}{\left\langle}
\newcommand{\ra}{\right\rangle}

\newcommand{\ct}[1]{\coth{\left(#1\right)}}

\newcommand{\eq}[2]{
\begin{equation}
#1
\label{#2}
\end{equation}
}

\begin{document}

\title{Nanoscale Heat Engine Beyond the Carnot Limit}

\author
{J. Ro{\ss}nagel}

\affiliation{Quantum, Institut f\"ur Physik, Universit\"at Mainz, D-55128 Mainz, Germany}

\author{O. Abah}
\affiliation{Institute for Theoretical Physics, University of Erlangen-N\"urnberg, D-91058 Erlangen, Germany}

\author{F. Schmidt-Kaler}
\affiliation{Quantum, Institut f\"ur Physik, Universit\"at Mainz, D-55128 Mainz, Germany}

\author{K. Singer}
\affiliation{Quantum, Institut f\"ur Physik, Universit\"at Mainz, D-55128 Mainz, Germany}

\author{E. Lutz}
\affiliation{Institute for Theoretical Physics, University of Erlangen-N\"urnberg, D-91058 Erlangen, Germany}

\date{\today}

\begin{abstract}
We consider a quantum Otto cycle for a time-dependent harmonic oscillator coupled to a squeezed thermal reservoir. We show that the efficiency at maximum power increases with the degree of squeezing, surpassing the standard Carnot limit and  approaching unity exponentially for large squeezing parameters. We further propose an experimental scheme to implement such a model system by using a single trapped ion in a linear Paul trap with special geometry.  Our analytical investigations are supported by Monte Carlo simulations that demonstrate the feasibility of our proposal.  For realistic trap parameters, an increase of the efficiency at maximum power of up to a factor of four is reached, largely exceeding the  Carnot bound.
\end{abstract}

\maketitle

Heat engines are important devices that convert heat into useful mechanical work. Standard heat engines run cyclically between two thermal (equilibrium) reservoirs at different temperatures $T_1$ and $T_2$. The second law of thermodynamics  restricts their efficiencies to the  Carnot limit, $\eta_c = 1-T_1/T_2$ ($T_1<T_2$) \cite{cen01}. Triggered by the pioneering study of Scovil and Schulz-DuBois on maser heat engines~\cite{Sco59} and boosted by the advances in nanofabrication, an intense theoretical effort has been devoted to the investigation of their properties in the quantum regime, see e.g. Refs.~\cite{ali79,kos84,gev92a,scu02,lin03,kie04,rez06,qua07,hen07a}.
In particular, theoretical studies have indicated that the efficiency of an engine may be increased beyond the standard Carnot bound by coupling it to an engineered (nonequilibrium) quantum coherent~\cite{scu03} or quantum correlated~\cite{dil09} reservoir (see also the related Refs.~\cite{scu10a,scu11,dor13,nal13} for photocell heat engines).
These  stationary nonthermal reservoirs are characterized by a temperature as well as additional parameters that quantify the degree of quantum coherence or quantum correlations. The maximum efficiency that can be reached in this nonequilibrium setting is limited by a generalized Carnot efficiency that can surpass the standard Carnot value~\cite{aba13}.
Quantum reservoir engineering techniques are powerful tools that enable the realization of arbitrary thermal and nonthermal environments~\cite{poy96}. Those techniques have first been experimentally demonstrated in ion traps~\cite{myat00}. Recently, they have been used to produce nonclassical states, such as entangled states, in superconducting qubits~\cite{mur12} and atomic ensembles~\cite{kra11}, as well as in circuit QED~\cite{sha13} and ion trap systems~\cite{lin13}.

In this Letter,  we develop a general theory of a quantum heat engine coupled to a squeezed thermal bath. We evaluate  both the efficiency and the efficiency at maximum power of the engine. We show that the {efficiency at maximum power} can be increased beyond the standard Carnot limit by exploiting the nonthermal properties of the reservoir, without questioning the universality of the general framework of thermodynamics. 
Squeezing is a general concept in quantum optics \cite{scu97}.
It may be characterized by a parameter $r$ such that the phase-space quadratures of a state are, respectively, multiplied by $e^{r}$ and $e^{-r}$~\cite{teich1989tutorial}.
Squeezed ground  states of the harmonic oscillator were first observed in photonic systems~\cite{slusher1985} and extensively studied in Ref.~\cite{breitenbach1997measurement}. Additionally, phononic \cite{mee96}, number state \cite{orzel01} and spin state \cite{esteve08} squeezing were  respectively observed  in ion systems and  Bose-Einstein condensates.
Squeezed states are important tools in high-precision spectroscopy~\cite{wineland1992spin}, quantum information~\cite{fur98}, quantum cryptography~\cite{hillery2000quantum}, and the detection of gravitational waves~\cite{cav81,abr92}. The properties of squeezed thermal states were theoretically examined in Refs.~\cite{fea88,kim89,vog91,wan93,mar93a,mar93b}.
The first experimental realization of squeezed thermal noise using a Josephson parametric amplifier was reported in Ref.~\cite{yur87}. However, the use of squeezed thermal baths  in quantum thermodynamics has been largely unexplored. In the following, we investigate an Otto cycle based on a time-dependent harmonic oscillator, a paradigm of quantum heat engines (see Refs.~\cite{scu02,lin03,kie04,rez06,qua07} and references therein). To analyze the effect of squeezing, we couple the engine to a high-temperature squeezed thermal reservoir, while the low-temperature reservoir is still purely thermal.
We find that the efficiency at maximum power  rises exponentially with the squeezing parameter $r$, surpassing the standard Carnot limit and converging towards unity exponentially. To illustrate our results, we apply our general formalism to a single-ion heat engine in a specially designed linear Paul trap coupled to laser reservoirs \cite{aba12}.
We further present for the first time a concrete experimental scheme to mimic the interaction with a squeezed thermal reservoir by combining reservoir and state engineering methods \cite{leibfried2003quantum}. Monte Carlo simulations with realistic trap parameters and laser interaction demonstrate the experimental realizability of such a scheme with current technology.  We show that the single-ion engine can run at maximum power up to an efficiency which is four times larger than the efficiency obtained with two thermal reservoirs and a factor of two above the standard Carnot bound.

\begin{figure}
\includegraphics[width=\columnwidth]{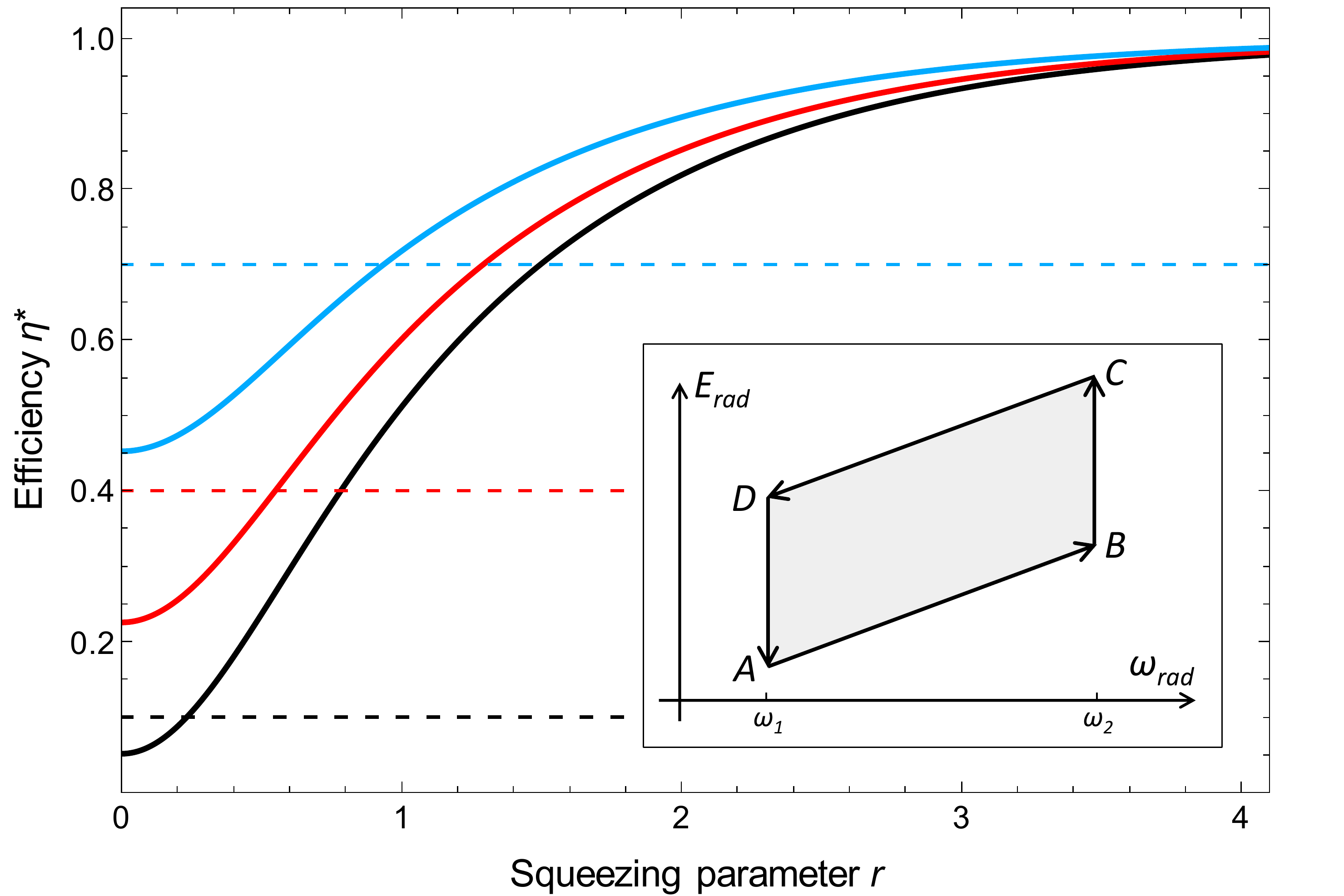}
\caption{Efficiency at maximum power~$\eta^\ast$, Eq.~(6), of the Otto engine plotted as a function of the squeezing parameter $r$. Black, red and blue lines (bottom to top) correspond to the temperature ratio $\beta_2/\beta_1 = 0.9, 0.6, \textrm{and}\, 0.3$, respectively. The dashed lines in the corresponding color denote the standard Carnot efficiency for each temperature ratio. The inset shows the energy-frequency diagram of an idealized Otto cycle. Squeezing is applied during the hot bath interaction between points $B$ and $C$,  leading to an increase of $\eta^{*}$  with $r$, approaching unity exponentially.}
\label{fig1}
\end{figure}

\paragraph{Otto engine with squeezed reservoir.}
We consider a quantum Otto cycle for a time-dependent harmonic oscillator that consists of four consecutive steps (expansion-heating-compression-cooling) \cite{scu02,lin03,kie04,rez06,qua07},  as shown in the inset of Fig.~1. During the expansion and compression phases, the frequency of the oscillator is modulated between $\omega_1$ and $\omega_2>\omega_1$. Heating and cooling result from the coupling to two heat baths at inverse temperatures $\beta_i=1/({ k_B} T_i)$, $(\beta_1>\beta_2)$, where ${ k_B}$ is the Boltzmann constant and $T_i$ the corresponding temperature.
The cycle starts in a thermal state $A$, at $\omega_1$ and at cold inverse temperature $\beta_1$,  with an average energy,
\eq{\la H\ra_A = \frac{\hbar\omega_1}{2}\,\ct{\frac{\beta_1\,\hbar\omega_1}{2}}.}{Ha}
During the initial isentropic compression from $A$ to $B$, the frequency increases  from $\omega_1$ to $\omega_2$. This transformation is unitary for an isolated system and the von~Neumann entropy is constant. The mean energy at point $B$ can be calculated by solving the Schr\"odinger equation for the driven quantum oscillator and is given by \cite{def08,def10},
\eq{\la H\ra_B = \frac{\hbar\omega_2}{2}Q^\ast_1\,\ct{\frac{\beta_1\,\hbar\omega_1}{2}},}{Hb}
where the parameter $Q^\ast_1$ characterizes the speed of the transformation.
The system is then coupled to a squeezed thermal reservoir at hot inverse temperature $\beta_2$ and squeezing parameter $r$, and, as a result, relaxes to a nondisplaced squeezed thermal state with mean phonon number, ${\la n(\beta_2,r)\ra = \la n\ra +(2 \la n\ra+ 1)\sinh^2(r)}$~\cite{mar93a}, where $\la n\ra = [\exp(\hbar \beta_2\omega_2)-1]^{-1}$ is the thermal  occupation number.
We assume the duration of this interaction to be much shorter than the duration of the isentropic process, and thus the frequency stays constant (corresponding to an isochoric process). The mean energy  at point $C$, $\la H\ra_C = \hbar\omega_2 \left<n(\beta_2,r)\right>$, is increased to \cite{mar93a},
\eq{\la H\ra_C =\frac{\hbar\omega_2}{2}\,\ct{\frac{\beta_2\,\hbar\omega_2}{2}}\Delta H(r),}{Hc}
where $\Delta H(r) = 1+(2+1/\la n\ra) \sinh^2 r$. In the following, we will keep the inverse temperature $\beta_2$ constant and vary the amount of squeezing $r$, hence the energy of the state $C$ (see Fig.~2).  During the following isentropic expansion, the frequency is brought back to its initial value $\omega_1$, and the mean energy at point $D$ reads,
\eq{\la H\ra_D=\frac{\hbar\omega_1}{2}Q^\ast_2\,\ct{\frac{\beta_2\,\hbar\omega_2}{2}}\Delta H(r).}{Hd}
The cycle is closed by coupling the system to the cold thermal bath. Because of the stochastic nature of this process (which is again isochoric), it destroys any phase relation and thus thermalizes the squeezed state.
 We stress that the above expressions are valid for arbitrary frequency modulations: ${Q^\ast_i = 1}$ for adiabatic and ${Q^\ast_i > 1}$ for nonadiabatic compression/expansion~\cite{hus53,def08,def10}.

Work is done by the oscillator during the compression and expansion phases, whereas heat is exchanged with the reservoirs during the thermalization steps. These quantities can be computed using Eqs.~(1)-(4) by evaluating the energy differences during each individual stroke \cite{aba12}. The efficiency $\eta$ of the engine is defined as the ratio of the net work produced  per cycle to the energy absorbed from the hot reservoir. Using Eqs.~(1)-(4), we find,
\eq{
\eta = 1 - \frac{\omega_1}{\omega_2}   \frac{\ct{\frac{\beta_1\hbar\omega_1}{2}} - Q^\ast_2 \ct{\frac{\beta_2\hbar\omega_2}{2}}\Delta H(r)}{ Q^\ast_1 \ct{\frac{\beta_1\hbar\omega_1}{2}} - \ct{\frac{\beta_2\hbar\omega_2}{2}}\Delta H(r)}.}{effi}
Expression \eqref{effi} is exact and valid for any temperature, squeezing, and adiabatic or nonadiabatic frequency modulation \cite{aba12}. In the remainder, we will focus on adiabatic modulation, $Q^\ast_i = 1$, which lead to the highest efficiencies. Note that the efficiency (5) is unaffected by the squeezing in this situation.
This is not the case for the efficiency at maximum power which we determine next.

Since the power of an engine vanishes at maximum efficiency, the efficiency at maximum power is the quantity of prime interest for practical applications \cite{cen01}. We maximize
the power, given by the  work produced by the engine divided by the cycle time,  with respect to the frequency difference $\Delta\omega=\omega_2-\omega_1$; we keep all other parameters, such as inverse temperatures $\beta_{1,2}$, squeezing parameter $r$, cycle time $\tau$ and the initial frequency $\omega_1$, constant.
In the high-temperature limit, $\hbar\beta_i\omega_i\ll 1$, we find that power is maximum when the frequencies satisfy the condition: ${\omega_2/\omega_1 = \sqrt{\beta_1(1+2\sinh^2{r})/\beta_2}}$. As a result, the efficiency at maximum power $\eta^\ast$ for adiabatic compression/expansion is given by,
\eq{\eta^\ast = 1 -\sqrt{\frac{\beta_2}{\beta_1(1+2\sinh^2{r})}},}{Eff_max_pow}
an expression which depends explicitly on the degree of squeezing.
For  thermal reservoirs ($r =0$), we recover the Curzon-Ahlborn efficiency, $\eta_{CA} = 1 -\sqrt{\beta_2/\beta_1}$~\cite{cur75}. Remarkably, the efficiency at maximum power $\eta^\ast$ rises with increasing squeezing $r$; it approaches unity exponentially for large squeezing parameter ($r\gg1$),
\eq{\eta^\ast \simeq 1 - \sqrt{2\, \frac{\beta_2}{\beta_1} \exp(-2r)}.}{eff_large}
Figure~\ref{fig1} shows the enhancement of the efficiency at maximum power with increasing squeezing  $r$ for different temperature ratios. While the Curzon-Ahlborn efficiency $\eta_{CA}$ is smaller than the Carnot limit (indicated  by dashed lines), we observe that $\eta^\ast$ may surpass it  even at moderate squeezing values. However, it does not exceed the generalized Carnot efficiency \cite{aba13, huang2012} (see Fig. 3),
\eq{
\eta_C^{gen} = 1-\frac{\beta_2}{\beta_1(1+2 \sinh^2(r))},
}{genCarnot}
which follows from  the second law  of thermodynamics applied to this  nonequilibrium situation. The latter  can be understood by noting that the standard Carnot efficiency is an expression of the second law  for one particular nonequilibrium configuration: two thermal reservoirs at two different temperatures. Equation (8) extends this result to a more general nonequilibrium setting that involves one thermal and one nonthermal reservoir.

\begin{figure}
\centering
\includegraphics[width=\columnwidth]{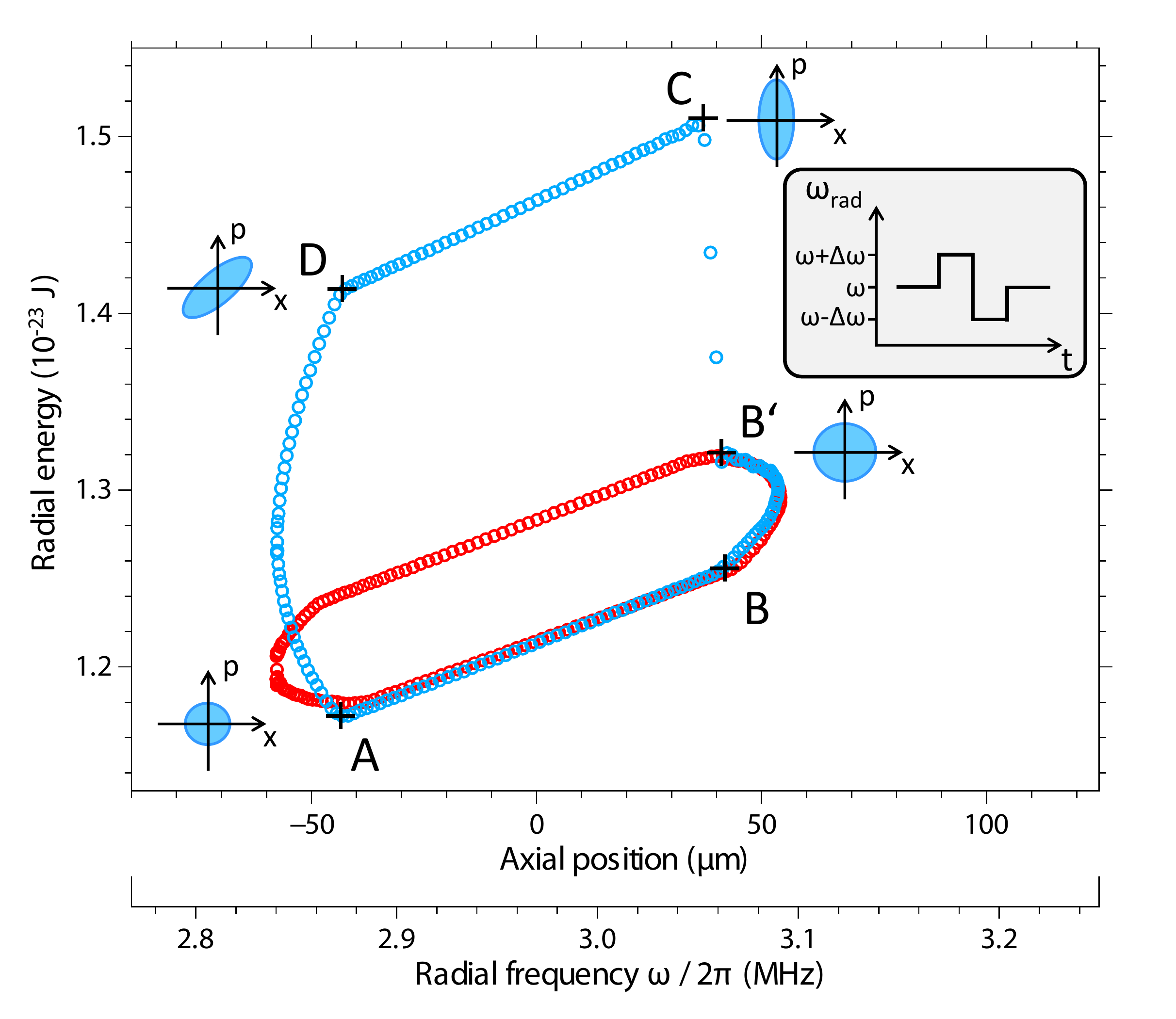}
\caption{Comparison of a thermal Otto cycle of the single ion heat engine (red, bottom) and a cycle with a hot squeezed thermal state (blue). The corresponding phase-space distributions of the ion are sketched next to points A,B',C and D, showing the change in temperature of a thermal state (rotational symmetric Gaussian distribution) and the squeezing of those states (ellipses). The interactions with the hot thermal bath ($BB'$) and the squeezing operation ($B'C$) are performed sequentially to discriminate both effects in their dynamics. The squeezing leads to a significant increase of the produced work (area of the enclosed region). Inset: Squeezing operation applied between points $B'$ and $C$. The radial trap frequency $\omega_\text{rad}$ is parametrically switched to higher and lower values.}
\label{fig2:cycle}
\end{figure}

\begin{figure}
\centering
\includegraphics[width=\columnwidth]{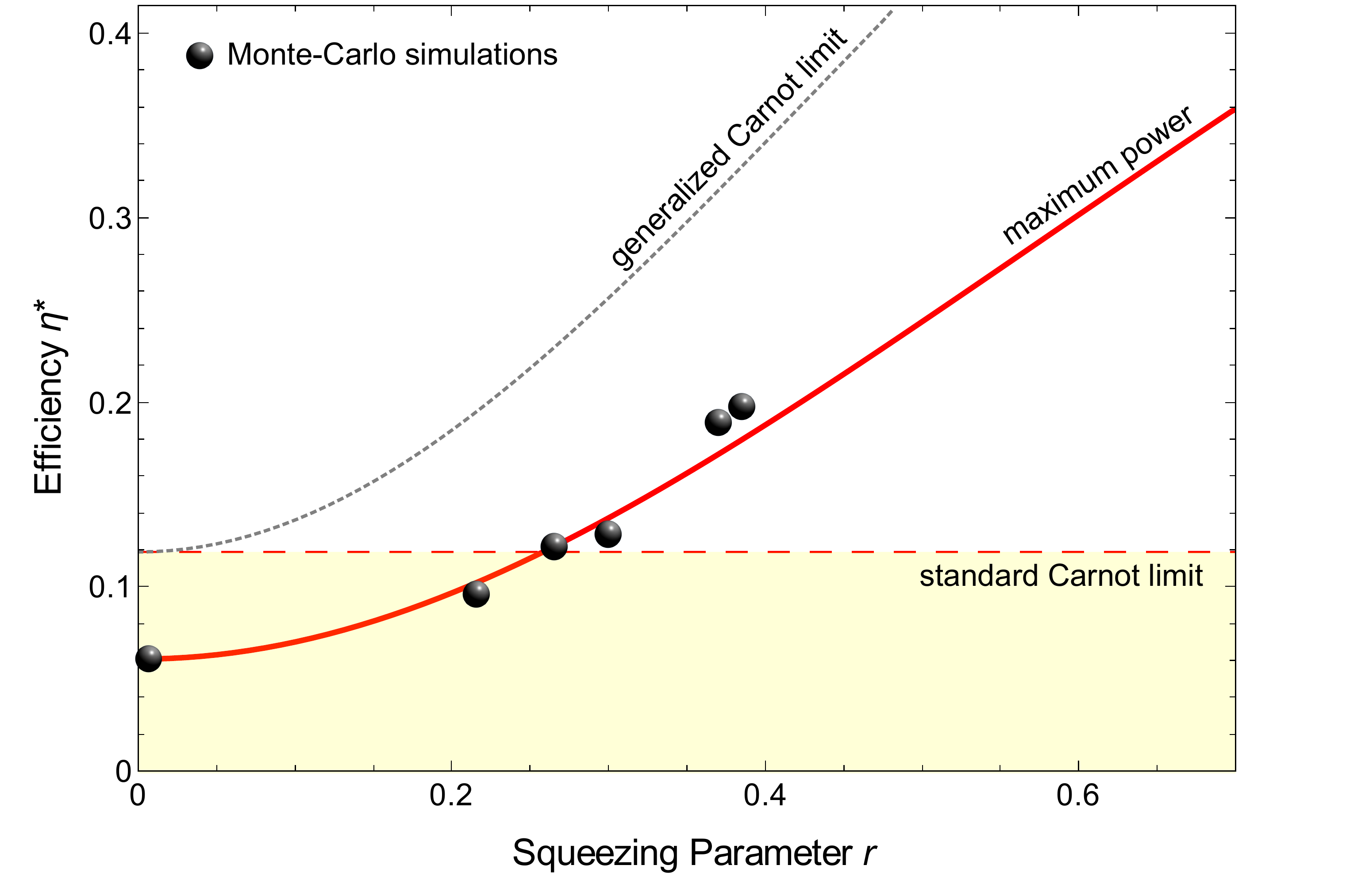}
\caption{Efficiency at maximum power $\eta^*$ given by the generalized Curzon-Ahlborn efficiency (\ref{Eff_max_pow}) as a function of the squeezing parameter $r$ (red line). The region below the red dashed line corresponds to all possible efficiencies in agreement with the standard Carnot limit. The results of the Monte-Carlo simulations (black dots) demonstrate within the given trap geometry that by squeezing the thermal state with $r\leq0.4$, the efficiency can be  increased by a factor of four, which is two times higher than the corresponding Carnot bound. The black dotted line shows the generalized Carnot limit (8) for an engine interacting with a hot squeezed thermal bath. The results shown are performed at a temperature ratio of $\beta_2/\beta_1=0.88$.}
\label{fig3:sim}
\end{figure}

\paragraph{Numerical simulations.}
To support our analytical findings, we consider the realistic proposal for an Otto heat engine presented in Ref.~\cite{aba12}.  This engine consists of a single ion confined in a  linear Paul trap and coupled to laser reservoirs. In contrast to conventional Paul traps, the radiofrequency electrodes that create the confining potential are tilted towards the trap axis with an angle $\theta$. Due to this geometry, the axial frequency  $\omega_{ax}$ of the ion is fixed, while the radial frequency $\omega_{rad}(z)$ is a function of the axial position $z$. The latter property is used to implement compression and expansion phases in the radial direction, while the ion moves back and forth along the trap axis. On the other hand, the work generated by the engine is stored in the axial oscillation.
We have performed semiclassical Monte-Carlo simulations of the Otto cycle using a partitioned Runge-Kutta integrator~\cite{singer2010colloquium}, including laser interaction (we  used the  typical realistic values: $\omega_{rad}\sim 3(2\pi)\,$MHz,  $\omega_{ax}\sim36 (2 \pi)\,$kHz and $\theta = 20^\circ$).
The dynamics of the ion's phase-space distribution are obtained by simulating an ensemble of several hundreds of classical trajectories, which follow the expected thermal phase distribution caused by the stochastic nature of photon scattering. To drive the heat engine, the ion is coupled alternatingly to a hot and a cold heat bath, realized by velocity dependent scattering forces of laser beams with different detuning.  An example of a cycle obtained with two thermal reservoirs is shown
in Fig. 2 (red cycle), as a function of the axial frequency of the ion and its radial position (which is experimentally accessible).
In the presence of a hot squeezed thermal bath, the state of the ion equilibrates to the temperature of the bath, but is additionally squeezed during the interaction \cite{vog91}. We mimic the coupling to such a squeezed thermal reservoir by combining reservoir engineering  (for the thermal component) and state engineering (for the squeezed component) \cite{com}. Squeezing of the state of the ion is implemented by modulating the radial confining potential at double the trap frequency~\cite{jan92,gal09a}. To ensure that an increase of efficiency can only be attributed to the squeezed state, the squeezing operation  is performed in such a manner that the mean value of the potential energy is not affected.
To this aim, the radial trap frequency $\omega_{rad}$ is first increased to $\omega'_{rad}=\omega_{rad}+\Delta \omega$ for a quarter of a radial oscillation period. Then the frequency is lowered to $\omega''_{rad}=\omega_{rad}-\Delta \omega$ for a quarter of a radial oscillation period, before it is returned to its initial value $\omega_{rad}$ (see inset in Fig.~\ref{fig2:cycle}). Since the total frequency change is zero, no work is performed by the squeezing operation  \cite{rem}. We numerically simulate engine cycles with different $\Delta \omega$ to achieve different squeezing parameters $r$ of the state of the ion. In order to  analyze the influence of the thermal  and the nonthermal part of the reservoir interaction separately, we  simulated the  coupling to the squeezed thermal bath in  two consecutive steps: first by heating  the state of the ion, followed by a squeezing operation.
This leads to cyclic processes as shown in Fig.~\ref{fig2:cycle}, where a cycle including a squeezed thermal bath (blue cycle) is compared to that employing a thermal bath with the same temperature ratio (red cycle), demonstrating the increase in energy due to the squeezing operation. Figure~\ref{fig3:sim} shows the resulting efficiencies, computed with Eq.~(5) with frequencies obeying the  optimality condition, for different squeezing parameters at a ratio of the bath temperatures of $\beta_1/\beta_2=0.88$: good agreement with the theoretical prediction for maximum power \eqref{Eff_max_pow} is attained.
We observe that for a squeezing parameter of 0.4 the efficiency is quadrupled, while reaching an efficiency two times higher than the standard Carnot limit. To achieve comparable values with a thermal bath interaction, while maintaining a maximized power output, an increase of the temperature ratio by $70\,\%$ would be needed, while having a power output still $35\,\%$ lower than the engine employing squeezing.

\textit{Concept of the experimental realization.} Let us now describe how the squeezed thermal state simulated above could be realized experimentally in an ion trap. Thermal reservoirs at different temperatures can be engineered via Doppler cooling, the Doppler temperature being  adjusted by employing electromagnetically induced transparency cooling and changing the lineshape through tuning of the laser parameters~\cite{morigi2000ground, schmidt2001laser}. Tailored electrical noise on the trap electrodes  can also be employed to efficiently heat the ion~\cite{turchette2000heating}.
Squeezing of the ground-state wave function of a trapped ion was first demonstrated using resolved sideband excitation on the second motional sideband~\cite{mee96}. However, this approach requires a system initially in the ground state and long interaction times.
As described above, another way to realize squeezing is to change suddenly the harmonic potential at double the trap frequency~\cite{jan92,gal09a}. This leads to an elliptical deformation of the phase space distribution of the thermal state (see Fig.~\ref{fig2:cycle}) and thus squeezes the state of the ion.

To our knowledge, such a scheme to achieve squeezed states has never been implemented experimentally.
We propose to make use of the tapered geometry of our setup, as it allows to change the radial confinement by shuttling the ion in axial direction. Thus, to squeeze the radial state of the ion, the latter can be driven along the trap axis at double the radial trap frequency. Recent studies have shown that fast transport of an ion along the trap axis on a sub-$\mu$s timescale is possible without additional heating~\cite{walther2012controlling}.
As the frequency of this modulation is two orders of magnitude higher than the axial resonant frequency, the two oscillations can be easily separated.
The proposed procedure avoids the use of an additional static potential to change the radial confining, as it could lead to parametric excitation of coherent oscillations. Those oscillations would indeed hide the signature of the squeezed state and perturb the experimental sequence for the heat engine cycle.

In order to run at maximum power, the engine should obey the optimality condition relating the frequencies of the oscillator to the  temperatures of the reservoirs. The maximum ratio~$\omega_2/\omega_1$ is limited by the opening angle~$\theta$ of the funnel shaped potential and by a maximum amplitude $a$ for the axial coherent oscillation. Considering realistic trapping potentials, this amplitude is chosen to be smaller than $a<1\,$mm~\cite{vahala2009phonon}.
These constraints limit the achievable squeezing parameters to~$r<0.6$ at maximum power. For the characterization of the resulting squeezed states, we may employ side-band spectroscopy and Raman transitions with standing light fields~\cite{mee96,Zeng1995,cirac1993dark}. The efficiency can be obtained from measuring the vibrational energy in the axial mode.

\textit{Conclusions.} We have shown that the efficiency at maximum power of a quantum Otto engine can be dramatically enhanced by coupling it to a squeezed thermal reservoir. While standard heat engines interact with thermal baths which are only characterized by their respective temperatures, the use of  nonthermal baths offers more degrees of control and manipulation, such as the amount of squeezing, that can be exploited  to increase the work produced. Our findings pave the way for a first experimental demonstration of the usefulness of reservoir and state engineering techniques in quantum thermodynamics and the realization of more efficient nano-engines.

\begin{acknowledgments}
We thank C.T.~Schmiegelow for comments and acknowledge support by the Volkswagen-Stiftung, the DFG-Forschergruppe (FOR 1493), the EU-project DIAMANT (FP7-ICT), the DFG (contract No LU1382/4-1) and the COST action MP 1209 "Thermodynamics in the quantum regime".
\end{acknowledgments}

\bibliography{bibliography}

\begin{thebibliography}{62}
\bibitem [{com()}]{com}%
  \BibitemOpen
  \href@noop {} {}\bibinfo {note} {The resulting squeezed thermal state of the
  ion is identical to that obtained from the coupling to a squeezed thermal
  reservoir. However, the relaxation to this state follows the dynamics imposed
  by a thermal reservoir. The latter has no effect on the evaluation of the
  efficiency of the engine, since relaxation is assumed to be negligibly short.
  Only the relaxed state matters.}\BibitemShut {Stop}%
\bibitem [{rem()}]{rem}%
  \BibitemOpen
  \href@noop {} {}\bibinfo {note} {The small radial frequency change seen in
  Fig. 2 is due to the axial motion $z(t)$ of the ion during the squeezing, and
  not to the squeezing itself. A similar effect is observed during the
  interaction with the thermal bath BB'.}\BibitemShut {Stop}%
\end{thebibliography}

\end{document}